\documentclass[aps,prl, twocolumn,floatfix,superscriptaddress,preprintnumbers,tightenlines,showpacs,showkeys,nofootinbib,notoccite]{revtex4-1}
\usepackage[utf8]{inputenc}
\usepackage[justification=justified,singlelinecheck=false]{caption}
\usepackage[colorlinks=true,citecolor=blue,linkcolor=blue]{hyperref}
\usepackage[normalem]{ulem}
\usepackage{amsmath,amssymb, mathrsfs}
\usepackage{epsfig}
\usepackage{graphicx}               % Standard graphics package
\usepackage{url}
\usepackage{color}
\usepackage{slashed}
\usepackage{multirow}
\usepackage{placeins}
\usepackage[dvipsnames]{xcolor}
\usepackage{epstopdf}
\usepackage{soul}
\usepackage{tikz}
\usepackage[capitalise, english]{cleveref}
\usepackage{siunitx}
\usepackage{xspace}
\usepackage{booktabs}
\usetikzlibrary{trees}
\usetikzlibrary{decorations.pathmorphing}
\usetikzlibrary{decorations.markings}
\usepackage{subfig}

\newcommand\myshade{80}
\colorlet{mylinkcolor}{ForestGreen}
\colorlet{mycitecolor}{Red}
\colorlet{myurlcolor}{violet}

\hypersetup{
  linkcolor  = mylinkcolor!\myshade!black,
  citecolor  = mycitecolor!\myshade!black,
  urlcolor   = myurlcolor!\myshade!black,
  colorlinks = true
}
\definecolor{jblue}{RGB}{20,50,100}
\definecolor{npurple}{RGB} {153, 51, 204}
\definecolor{wred}{RGB}{217,0,56}
\definecolor{white}{RGB}{255,255,255}

\definecolor{korange}{RGB}{235, 80,  43}
\definecolor{korange2}{RGB}{245, 100,  63}
\definecolor{kyelloworange}{RGB}{255, 210,  110}
\definecolor{kyelloworange2}{RGB}{240, 170,  90}
\definecolor{kred}{RGB}{204,  102, 153}
\definecolor{kpurple}{RGB}{153,  61, 190}
\definecolor{kpurplelight}{RGB}{213,  161, 230}

 \definecolor{tobycolour}{rgb}{.5,.0,.5}

\DeclareSIUnit\year{yr}
\DeclareSIUnit\pc{pc}
\DeclareSIUnit\ergs{ergs}
\DeclareSIUnit\msun{\ensuremath{M_\odot}}
\allowdisplaybreaks

\newcommand{\ev}[1]{\ensuremath{\left\langle #1 %
                     \right\rangle}} % Expectation value
\makeatletter
\providecommand*{\diff}%
  {\@ifnextchar^{\DIfF}{\DIfF^{}}}
\def\DIfF^#1{%
  \mathop{\mathrm{\mathstrut d}}%
    \nolimits^{#1}\gobblespace}
\def\gobblespace{%
  \futurelet\diffarg\opspace}
\def\opspace{%
  \let\DiffSpace\!%
  \ifx\diffarg(%
    \let\DiffSpace\relax
  \else
    \ifx\diffarg[%
      \let\DiffSpace\relax
    \else
        \ifx\diffarg\{%
        \let\DiffSpace\relax
      \fi\fi\fi\DiffSpace}

\usepackage{tikz,xcolor,hyperref}

\definecolor{lime}{HTML}{A6CE39}
\DeclareRobustCommand{\orcidicon}{\hspace{-1mm}
	\begin{tikzpicture}
	\draw[lime, fill=lime] (0,0) 
	circle [radius=0.16] 
	node[white] {{\fontfamily{qag}\selectfont \tiny \,ID}};
	\draw[white, fill=white] (-0.0525,0.095) 
	circle [radius=0.007];
	\end{tikzpicture}
	\hspace{-3mm}
}

\foreach \x in {A, ..., Z}{\expandafter\xdef\csname orcid\x\endcsname{\noexpand\href{https://orcid.org/\csname orcidauthor\x\endcsname}
			{\noexpand\orcidicon}}
}

\usepackage{xcolor}      


\begin{document}
\preprint{HRI-RECAPP-2023-03}

%%=============================================================================
\title{Neutrinos from the Sun can discover dark matter-electron scattering}

\author{Tarak Nath Maity\orcidA{}}
\email{tarak.maity.physics@gmail.com}
\affiliation{School of Physics, The University of Sydney and ARC Centre of Excellence for Dark Matter Particle Physics, NSW 2006, Australia}
\affiliation{Harish-Chandra Research Institute, A CI of Homi Bhabha National Institute, Chhatnag Road, Jhunsi, Prayagraj (Allahabad) 211019, India}
\affiliation{Regional Centre for Accelerator-based Particle Physics, Harish-Chandra Research Institute, Prayagraj (Allahabad) 211019, India}
\affiliation{Centre for High Energy Physics, Indian Institute of Science, C.\,V.\,Raman Avenue, Bengaluru 560012, India}
\author{Akash Kumar Saha\orcidB{}} 
\email{akashks@iisc.ac.in}
\affiliation{Centre for High Energy Physics, Indian Institute of Science, C.\,V.\,Raman Avenue, Bengaluru 560012, India}
\author{Sagnik Mondal\orcidC{}} 
\email{sagnikmondal@iisc.ac.in}
\email{mondalsagnikmandal@gmail.com}
\affiliation{Centre for High Energy Physics, Indian Institute of Science, C.\,V.\,Raman Avenue, Bengaluru 560012, India}

\author{Ranjan Laha\orcidD{}} 
\email{ranjanlaha@iisc.ac.in}
\affiliation{Centre for High Energy Physics, Indian Institute of Science, C.\,V.\,Raman Avenue, Bengaluru 560012, India}

\date{\today}

%%%%%%%%%%%%%%%%%%%%%%%%%%%%%%%%%%%%%%%%%%%%%%%%%%%%%%%%%%%%%%%%%%%

\begin{abstract}
We probe dark matter-electron scattering using high-energy neutrino observations from the Sun. Dark matter (DM) interacting with electrons can get captured inside the Sun. These captured DM may annihilate to produce different Standard Model (SM) particles. Neutrinos produced from these SM states can be observed in IceCube and DeepCore. Although there is no excess of neutrinos in the solar direction, we find that the current datasets of IceCube and DeepCore set the strongest constraint on the DM-electron scattering cross section in the DM mass range $10$\,GeV to $10^5$\,GeV. Therefore our work implies that future observations of the Sun by neutrino telescopes have the potential to discover the DM-electron interaction.
\end{abstract}

\maketitle

\section{Introduction}
\label{sec:introduction}  
%%%%%
Have we searched for all possible ways in which new physics can manifest itself in our existing datasets?  This is an important question to consider in beyond the SM searches.  The effects of new physics may be present in existing datasets, but we will be able to find them only if we use the right observables to interpret the data-sets. The above-mentioned strategy may be a discovery probe of new physics. We show that by using an existing IceCube dataset, we can set the strongest constraint on dark matter (DM)-electron scattering cross sections. This implies that by utilizing similar near-future datasets, it is possible to discover nongravitational interactions of DM.

DM is a ubiquitous component of the Universe: this is an inescapable conclusion from various cosmological and astrophysical observations\,\cite{Planck:2018vyg, Green:2021jrr, Bertone:2004pz, Strigari:2012acq}.  These observations imply the presence of DM through its gravitational interactions.   In spite of this large body of evidence, we neither know the DM candidate nor its nongravitational interactions with various SM particles\,\cite{Buchmueller:2017qhf, Ahlers:2018mkf, Krnjaic:2022ozp, Batell:2022dpx, PerezdelosHeros:2020qyt, Slatyer:2021qgc, Cooley:2021rws, Kahn:2021ttr}.  Since new physics can be present in a variety of different ways, it is important to search for all different couplings of DM with SM particles.  In this work, we focus on the DM-electron scattering for leptophilic DM\,\cite{Krauss:2002px, Baltz:2002we, Ma:2006km, Hambye:2006zn, Bernabei:2007gr, Cirelli:2008pk, Chen:2008dh, Bi:2009md, Cao:2009yy, Goh:2009wg, Bi:2009uj, Ibarra:2009bm, Davoudiasl:2009dg,Dedes:2009bk, Kopp:2009et, Cohen:2009fz, Chun:2009zx, Chao:2010mp, Haba:2010ag,Ko:2010at, Carone:2011iw, Schmidt:2012yg, Das:2013jca, Dev:2013hka, Kopp:2014tsa, Chang:2014tea, Agrawal:2014ufa, Bell:2014tta, Freitas:2014jla,Cao:2014cda, Boucenna:2015tra,Lu:2016ups, Chauhan:2016joa, Garani:2017jcj, Duan:2017pkq, Duan:2017qwj,Chao:2017emq,Li:2017tmd, Ghorbani:2017cey,Sui:2017qra,Han:2017ars,  Madge:2018gfl,Junius:2019dci, Bell:2019pyc, YaserAyazi:2019psw, Joglekar:2019vzy,  Ghosh:2020fdc, Joglekar:2020liw, Chakraborti:2020zxt, Bell:2020lmm, Horigome:2021qof, Garani:2021ysl,Nguyen:2025ygc}. 

The search for DM-electron couplings has made rapid progress over the past decade.  Various techniques have probed large regions of the DM-electron coupling parameter space\,\cite{Essig:2011nj, Essig:2015cda, XENON:2019gfn, SuperCDMS:2018mne, DarkSide:2018ppu, DarkSide-50:2022hin, DAMIC:2019dcn, EDELWEISS:2020fxc, Crisler:2018gci,SENSEI:2020dpa, Dror:2020czw, Maity:2020wic, PandaX-II:2021nsg, Maity:2022enp, Bhowmick:2022zkj, Bardhan:2022bdg, DAMIC-M:2023gxo,Wadekar:2019mpc, Ali-Haimoud:2021lka, Nguyen:2021cnb, Buen-Abad:2021mvc, Giovanetti:2021izc, Bose:2021cou, Bellomo:2022qbx, Chatterjee:2022gbo}.  These searches motivate us to ask: is there an observable that can discover DM-electron scattering cross section beyond what has already been probed? In this work, we obtain the most stringent constraint on the DM-electron scattering cross-section, $\sigma_e$, for a wide range of DM masses using current IceCube datasets.

%%%%%
\begin{figure}
\centering
%\captionsetup{width=\linewidth}
\includegraphics[width=\columnwidth]{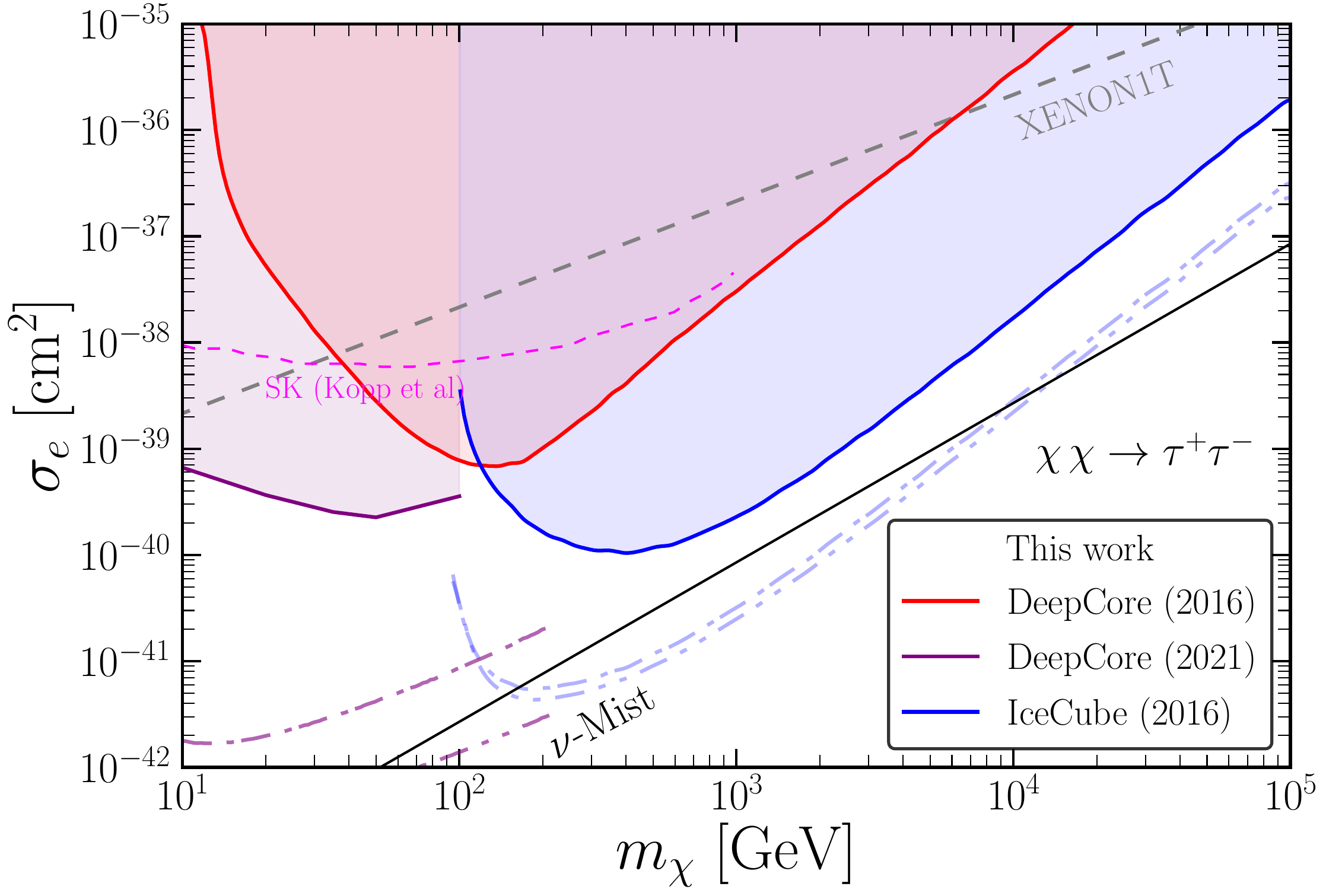}
\caption{Excluded region of DM-electron scattering cross section, $\sigma_e$, for DM annihilating to $\tau^+ \tau^-$. The purple, red, and blue shaded regions are excluded by DeepCore (2021), DeepCore (2016), and IceCube (2016) data respectively. The XENON1T and SK bounds are shown by the dashed gray and magenta lines respectively. Along the dot-dashed, double dot-dashed light blue (purple) lines, DM induced neutrino events would be equal to the same of two different models for solar atmospheric neutrino background (SA$\nu$) events in IceCube (DeepCore). Above the black solid line DM thermalizes within the solar age, as discussed in the Appendix.} 
\label{fig:limit} 
\end{figure}

The Sun, as the nearest star to Earth, is an interesting target for DM.  The Sun has been moving through the Milky Way DM halo over its lifetime ($\gtrsim$ Gyr). If there is an interaction between the DM and SM states, then a fraction of local DM particles are captured within the Sun, for DM masses above $\sim 5\,$GeV. There are a large number of electrons inside the Sun, which would facilitate DM capture through electron scattering. These DM particles may annihilate into different SM particles (whose decay, hadronization, electroweak correction (EW) etc. produce neutrinos)\footnote{Unless specified, we do not distinguish between neutrinos and antineutrinos in this work.} including neutrinos. Neutrinos are the only SM particles that can escape from the Sun, and can be detected in Earth-based neutrino telescopes. Using existing results of IceCube and DeepCore, we set the strongest limits on $\sigma_e$ for the DM mass range $\sim 10$\,GeV$-100$\,TeV.

In Fig.\,\ref{fig:limit}, we show our derived limits for $\sigma_e$ as a function of the DM mass. Here we assume that the DM particles, captured within the Sun, annihilate to $\tau^+ \tau^-$. The solid blue, red, and purple lines show the bounds derived in this work using the IceCube and DeepCore data sets from Refs.\,\cite{IceCube:2016dgk,IceCube:2021xzo}. Our framework probes new regions of parameter space in the DM mass range $\sim 10$ GeV-$ 100\,$TeV. This suggests that solar observations using neutrino telescopes can potentially probe DM-electron interactions. The dot-dashed and double dot-dashed lines represent ``$\nu$-Mist" which we will discuss later. The black solid line represents the DM-electron scattering cross section above which DM thermalizes within the Sun.

%%%%%%%%%%%%%%%%%%%%%%%%
\section{DM capture in the Sun}
\label{sec:cap}
%%%%%
While the Sun is traversing the Milky Way halo, DM particles can get gravitationally attracted toward the Sun's potential. Further nongravitational interactions may lead to scattering between the DM and Solar constituents. After scattering, if the final velocity of the DM particle is less than the Solar escape velocity, then it gets trapped inside the Sun. Furthermore, these captured DM particles can thermalize within the Sun due to further scattering with electrons and be concentrated within $\sim 1\%$ of the solar radius \cite{Kopp:2009et,1987ApJ...321..560G,1987ApJ...321..571G,Garani:2017jcj, Baratella:2013fya,Zentner:2009is}. These captured DM can annihilate to produce primary and secondary neutrinos that can be detected by various neutrino telescopes. We adopt the model-independent approach of considering a nonzero DM-electron cross section leading to DM capture inside the Sun (see Refs.\,\cite{Silk:1985ax, Krauss:1985ks, Srednicki:1986vj, Ng:1986qt, Ritz:1987mh, Olive:1987av, Super-Kamiokande:2004pou, Cirelli:2005gh, Mena:2007ty, Savage:2008er, Hooper:2008cf, Agrawal:2008xz, Peter:2009mk, Koushiappas:2009ee, Zentner:2009is, Niro:2009mw, Ellis:2009ka, Barger:2011em, IceCube:2011aj, Bell:2012dk, Bernal:2012qh, Silverwood:2012tp, IceCube:2012ugg, Ibarra:2013eba, Choi:2013eda, Allahverdi:2014eca, Kumar:2015nja, Catena:2015iea, Super-Kamiokande:2015xms, ANTARES:2015vis, Allahverdi:2015ssa, Danninger:2014xza, Rott:2015nma, Feng:2016ijc, Lopes:2016ezf, Catena:2016ckl, Baum:2016oow, Allahverdi:2016fvl, Smolinsky:2017fvb,  Widmark:2017yvd, Fornengo:2017lax, Demidov:2017mmw, HAWC:2018szf, Rott:2019stu, Niblaeus:2019gjk, Nunez-Castineyra:2019odi, Brenner:2020mbp, Acevedo:2020gro, Bell:2021pyy, Bell:2021esh, Bose:2021yhz, Zakeri:2021cur, Bose:2022ola, Nguyen:2022zwb, Ray:2023auh} for similar studies with DM-nucleon scattering).

In the weak $\sigma_e$ limit, the capture rate of DM particles having mass, $m_{\chi}$\,\cite{1987ApJ...321..560G, 1987ApJ...321..571G}
\begin{eqnarray}
\label{eq:cweak}
 C_\odot &=& \int^{R_\odot}_0{4\pi r^2 dr}\int^\infty_0{du_{\chi}\left(\frac{\rho_{\chi}}{m_{\chi}}\right)}\frac{f_{v_\odot}\left(u_\chi\right)}{u_\chi}w(r) \nonumber\\
&&\int^{v_e(r)}_0R^-_e\left(w\rightarrow v\right)dv,
\end{eqnarray}
where $R_\odot$ is the solar radius and $\rho_\chi=0.3$ GeV/cm$^3$ is the local DM density. Here $f_{v_\odot}(u_\chi)$ is the DM velocity distribution in the solar rest frame. DM velocities at asymptotic and $r$ distance are $u_\chi$ and $w(r)=\sqrt{u_\chi^2+v_e^2(r)}$ respectively with escape velocity $v_e(r)$. The integrand in Eq.\,\eqref{eq:cweak}, $R^-_e\left(w\rightarrow v\right)$ is the differential scattering rate for DM particles with velocity $w$  scattering to a final velocity $v$, such that $v<w$. The form of $R^-_e\left(w\rightarrow v\right)$ for velocity-independent, isotropic DM–electron scattering, with Maxwell Boltzmann distribution for electrons inside the Sun, is given by\,\cite{1987ApJ...321..571G, Garani:2017jcj}
\begin{eqnarray}
\label{eq:Rminus}
R^-_e\left(w\rightarrow v\right) &=& \frac{2}{\sqrt{\pi}}\,\frac{\mu_+^2}{\mu} \,\frac{v}{w}\,n(r)\,\sigma_e \,[\,\chi(-\alpha_-,\alpha_+) \nonumber \\ 
&&+\chi(-\beta_-,\beta_+)e^{\frac{\mu(w^2-v^2)}{u_e(r)^2}}\,],
\end{eqnarray}
where,
\begin{eqnarray}
\label{eq:RminusDetails}
&& \mu=\frac{m_\chi}{m_e},~ \mu_\pm=\frac{\mu\pm 1}{2},~\chi(a,b)= \int_{a}^{b} e^{-y^2} \,dy\, \\
&& \alpha_\pm=\frac{\mu_+v\pm \mu_-w}{u_e(r)},~ \beta_\pm=\frac{\mu_-v\pm \mu_+w}{u_e(r)},  u_e(r)=\sqrt{\frac{2\,T_\odot (r)}{m_e}} \nonumber 
\end{eqnarray}
with $n(r)$, $T_\odot (r)$ being the number density and temperature of solar electrons as a function of $r$. Here we are using the AGSS09 solar model for our analysis\,\cite{Vinyoles:2016djt}. 

Some types of DM-electron interactions generate DM-nuclear scattering at the loop level, which would give an additional contribution in the capture rate. Assuming a fermionic DM, $\chi$, interacting with a lepton, $l$, the general dimension-6 four-fermion interaction is given by
\begin{eqnarray}
\mathcal{L} = \sum_i G \left(\bar{\chi}\,\Gamma^i_\chi\,\chi \right) \left(\bar{\ell}\,\Gamma^i_\ell \,\ell \right) \,\,,
\end{eqnarray}
  where $G = 1/\Lambda^2$, with $\Lambda$ being the effective field theory cutoff. In the above expression, the index $i$ runs over all possible Lorentz structures $\Gamma^i_\chi$, $\Gamma^i_\ell$. It has been shown in Ref.\,\cite{Kopp:2009et} that for pseudoscalar and axialvector lepton currents, the loop-level DM-nucleon cross sections are zero at all orders. Besides, we assume that the DM-electron scattering is velocity independent. In such a case, Ref.\,\cite{Kopp:2009et} found that for axial-axial type DM-lepton coupling, the corresponding loop-level DM-nucleon cross section will be subdominant (see Table 1 of Ref.\,\cite{Kopp:2009et}). In our work, we assume that DM predominantly scatters with electrons and the loop-level DM-nucleon cross section is subdominant. Such models have been discussed in Refs.\,\cite{Kopp:2009et,Bell:2019pyc} (see  \cite{Garani:2021ysl} for the two-photon exchange calculation). In our work, we do not go into the details of such particle physics models but rather discuss the interesting phenomenological implications. A detailed study of such leptophilic DM models, assuming all possible spins and interactions of DM, is left for future work.

For DM mass $\gtrsim 10\,$GeV, and within our cross section range of interest, the capture rate is 
\begin{equation}
\label{eq:crateapprox}
C_\odot \approx 2.14 \times 10^{19} \left( \frac{10^3 \,{\rm GeV}}{m_{\chi}} \right)^2 \left( \frac{\sigma_e}{10^{-40} \,{\rm cm}^2}\right) \, {\rm s^{-1}}
\end{equation}

In the previous version of this work, in the capture rate we incorporated a kinematic condition (following Eq. 2.12
of Ref.\,\cite{1987ApJ...321..571G}) along with Eq.\,\eqref{eq:Rminus} that is strictly valid for a stationary target. Under this condition, our capture rate matches with Refs.\,\cite{Kopp:2009et, Garani:2017jcj}. After discussing with the authors of Ref.\,\cite{Nguyen:2025ygc}, we realized that this condition is not applicable for solar electrons, given their finite velocity ($\sim 0.07c$). This leads to a factor of $\sim$7 enhancement in the capture rate. Our results agree with Ref.\,\cite{Nguyen:2025ygc}. We thank the authors of Ref.\,\cite{Nguyen:2025ygc} for detailed discussions on the capture rate which uncovered this issue.

\section{Neutrino flux}
\label{sec:nuFlux}
Captured DM inside the Sun may annihilate into different SM final states. There is an interplay between the rate at which DM gets captured and annihilated away. Under the equilibrium condition between the capture and annihilation rate (true for the DM regions that we are probing), the latter is
\begin{equation}
\label{eq:annrate}
\Gamma_{\rm ann} \equiv \frac{C_\odot}{2}\,.
\end{equation}
Note that DM evaporation is ineffective in our mass range of interest ($\gtrsim 10$ GeV)\,\cite{Garani:2021feo}. We have also neglected the effect of model dependent DM self-interaction. The equilibrium timescale depends on the DM mass, DM-electron scattering, and DM annihilation cross sections. In the Appendix B, we show the regions where the equilibrium is achieved in thermally averaged DM annihilation cross sections and the DM-electron scattering cross section plane for various choices of DM masses.  

Only primary and secondary neutrinos of energies $\lesssim500$ GeV (due to neutrino attenuation, discussed later) will emerge out of the Sun without substantial attenuation. The differential neutrino energy flux at earth is
\begin{eqnarray}
E_{\nu}^2 \frac{d \phi_{\nu}}{d E_{\nu}} = \frac{\Gamma_{\rm ann}}{4 \, \pi \, D_{\odot}^2} \times  E_{\nu}^2 \, \frac{dN_{\nu}}{dE_{\nu}},
\label{eq:neutrino_flux}
\end{eqnarray} 
where $D_{\odot}$ is the Earth-Sun distance.\footnote{ Time variation of $D_{\odot}$ affects our results insignificantly. We have fixed it to May 1, since IceCube data are taken during austral winter.} The neutrino spectrum is denoted by $dN_{\nu}/dE_{\nu}$ per DM annihilation. The Sun's center is a dense environment; some particles (like muons) will interact before decay, and others (like top quark, tau) will decay before interaction. Therefore, neutrino spectra at the center of the Sun will be different from the typical Galactic halo environment spectrum. We have utilized the results from the publicly available  code, \texttt{\raisebox{0.82\depth}{$\chi$}aro$\nu$}\,\cite{Liu:2020ckq} (see also\,\cite{Blennow:2007tw, Niblaeus:2019gjk, Gondolo:2004sc, Bringmann:2018lay}), which includes these effects along with other important effects like EW corrections\,\cite{Bauer:2020jay} (for DM masses $>500$\,GeV).

 The propagation media of neutrinos from the solar center to the detector include the Sun, the vacuum, Earth's atmosphere, and Earth rock (negligible impact for this work). The primary propagation effects are the trapping of neutrinos and tau regeneration in the solar medium, and neutrino oscillation. Electron and muon neutrinos, once produced near the center of the Sun, mainly interact through charged current (CC) interactions and produce the corresponding charged leptons. These charged leptons thermalize within the Sun and thus essentially remove electron and muon neutrinos above the transparency energy (the energy below which neutrinos can escape the Sun). Tau leptons, produced from interactions of tau neutrinos, are so short lived that they will decay before scattering\,\cite{Liu:2020ckq,Albuquerque:2000rk}. Their decay products include tau neutrinos with energies less than ($\sim 20\%$ reduction) that of the initial tau neutrinos. For initial tau neutrinos with energies above the corresponding transparency energy, this process will continue until the final tau neutrino reaches energy below the transparency energy. Finally, one needs to consider the effect of neutrino oscillations. We have incorporated all these effects using \texttt{nuSQuIDS}\,\cite{ Arguelles:2021twb}. The differential neutrino spectra obtained from \texttt{nuSQuIDS} is then used in Eq.\,\eqref{eq:neutrino_flux} to calculate the differential neutrino flux at the detector.

\section{Analysis}
\label{sec:ana}
Neutrinos produced from DM capture through DM-electron scattering will have observable signatures in terrestrial neutrino telescopes like IceCube, Super-Kamiokande (SK), etc. Current data do not show excess events from the solar direction. We have analyzed the current IceCube and DeepCore data-set to constrain $\sigma_e$. In our neutrino energy range of interest, neutrino detection topologies include cascades and tracks. Cascades are produced by neutral current  interactions of all neutrino flavors and by CC interactions of electron and tau neutrinos. CC interactions of muon neutrinos produce muons that produce tracklike signatures. \\        
{\bf IceCube \& DeepCore (2016)\,\cite{IceCube:2016dgk}:} In this analysis, IceCube utilizes austral winter data from May 2011 to May 2014. During this period, muon neutrinos from the solar direction will produce upgoing muon tracks. This facilitates differentiating the signal from the large downgoing atmospheric muon background. The directionality of the upgoing tracklike events serves as a proxy for the direction of neutrinos. Atmospheric neutrinos from the solar direction are an irreducible background to this search.  This analysis focuses only on the (anti)muon tracks produced by interactions of muon (anti)neutrinos. For neutrinos with energies $\gtrsim 100$\,GeV (from DM annihilations), the full instrumented volume of IceCube contributes to the solar DM annihilation sensitivity, whereas DeepCore is sensitive only to neutrinos with energies $\lesssim 100$\,GeV.

In \,\cite{IceCube:2016dgk} event rates are given with respect to the cosine of the solar opening angle ($\theta_{\rm Sun}$), the angle subtended by reconstructed (anti)muon with respect to the solar direction. The differential event rate as a function of $\theta_{\rm Sun}$ is
\begin{eqnarray}
\label{eq:eventrateICDC}
\frac{dN_{\theta_{\rm Sun}}}{d\cos(\theta_{\rm Sun})}= 2 &T &\int_{E_\nu^{\rm min}}^{E_\nu^{\rm max}} A_{\rm eff}(E_\nu) \frac{d \phi_{\nu}}{d E_{\nu}}\, \nonumber \\  && \frac{1}{\sqrt{2 \pi } \sigma_{\theta}} e^{-\frac{\left(\cos(\theta_{\rm Sun})-1\right)^2 }{2 \sigma_{\theta}^2}} dE_{\nu},
\end{eqnarray}
where $T=532\,$days, and a factor of 2 is there since cosine is an even function of $\theta_{\rm Sun}$. The energy dependent effective area ($A_{\rm eff}(E_\nu)$), lower, and upper limits of integration $E_\nu^{\rm min}$ and $E_\nu^{\rm max}$ are adopted from Ref.\,\cite{IceCube:2016dgk}. The dispersion ($\sigma_{\theta}$) also depends on energy through
\begin{equation}
\sigma_{\theta} = \bigg |\frac{\sqrt{2}\left(1-\cos[\Delta \theta(E_{\nu})]\right)}{2\, {\rm InverseErf(0.5)}}\bigg |,
\label{eq:sigma}
\end{equation}
where the energy dependency of median angular resolution $\Delta \theta(E_{\nu})$ is extracted from Ref.\,\cite{IceCube:2016dgk}. InverseErf is the inverse error function. Since this analysis is done with tracklike signatures, only muon neutrinos and muon antineutrinos contribute to the differential flux given in Eq.\,\eqref{eq:eventrateICDC}.

\begin{figure}
%\begin{center}
\centering
\includegraphics[width=\columnwidth]{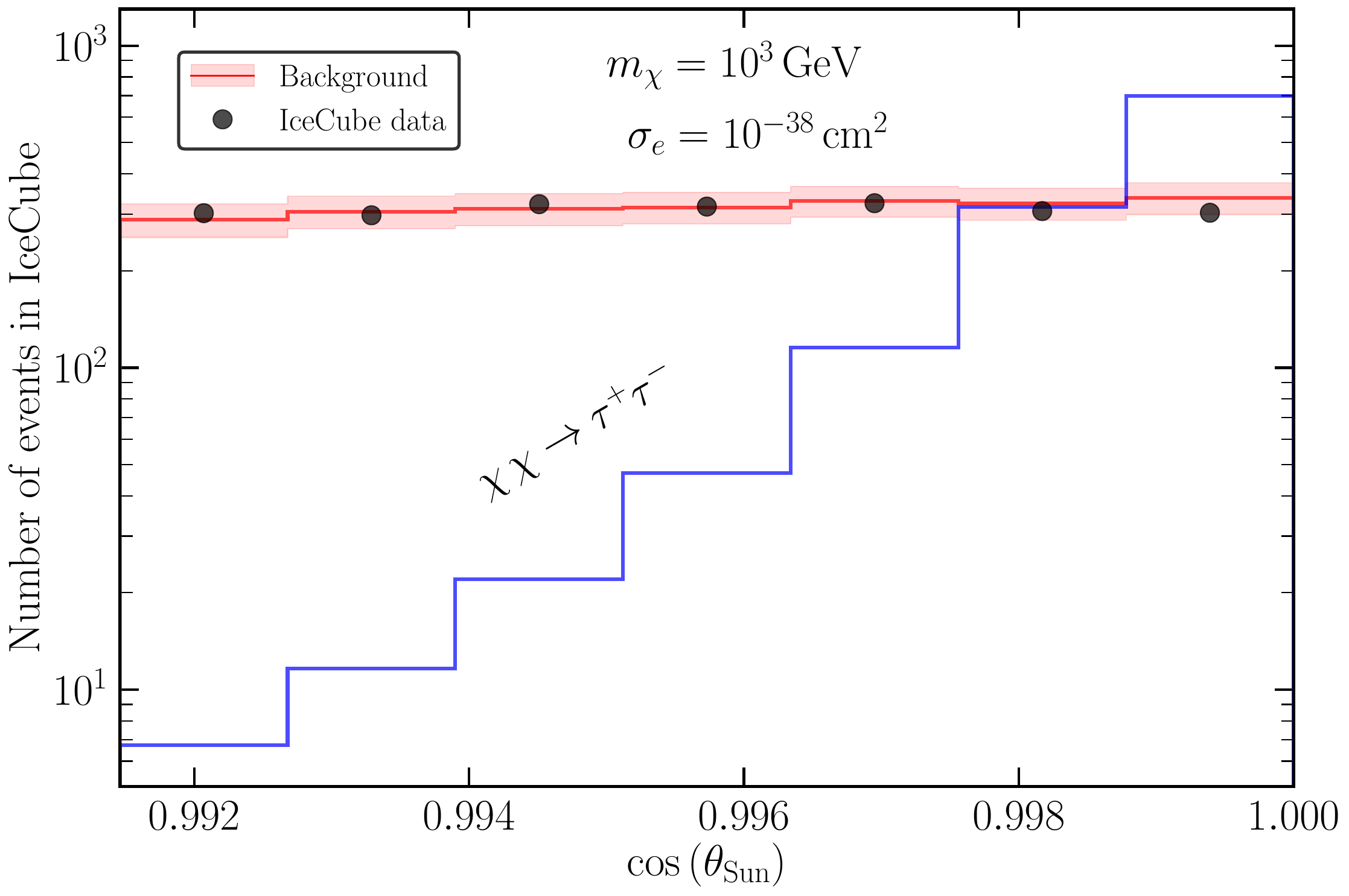}
\caption{Number of events against the $\cos(\theta_{\rm Sun})$ for captured DM annihilating to $\tau^+ \tau^-$ with $\sigma_e=10^{-38}\,{\rm cm}^2$ and $m_{\chi}=10^3\,$GeV is shown by blue solid line. Black dots, red solid line, and the shaded region show the data, expected atmospheric neutrino events, and $2\sigma$ uncertainty in the background events respectively. The calculation and the dataset are from IceCube (2016) analysis.}
%\end{center}
\label{fig:ang-dis} 
\end{figure}

Assuming DM is annihilating to $\tau^+\tau^-$, we show the IceCube event rate with the blue solid line in Fig.\,\ref{fig:ang-dis} for $m_{\chi}=1\,$TeV and $\sigma_e=10^{-38}\,{\rm cm}^2$ (this choice is consistent with other experiments). The black dots represent observed data points, and the red solid line with the shaded band depicts the background atmospheric neutrino events and its uncertainty. Clearly, the chosen DM parameter space is already ruled out by the current IceCube observation, showing the power of our technique. 

Numerically, we obtain the constraint on DM parameter space by performing a $\chi^2$ analysis. The $\chi^2$ estimator is given by
\begin{eqnarray}
\chi^2 \,=\, \sum_{i=1}^{7} \left(N^i_{\rm data} \,-\, N^i_{\rm bkg} \,-\, N^i_{\theta_{\rm Sun}} \right)^2/(\sigma^i)^2\,,\phantom{x}
\label{eq: chi-squraed}
\end{eqnarray}
where $N^i_{\rm data}$, $N^i_{\rm bkg}$, and $N^i_{\theta_{\rm Sun}}$ are the observed data (black points in Fig.\,\ref{fig:ang-dis}), atmospheric neutrino events (solid red line in Fig.\,\ref{fig:ang-dis}), and signal events (by integrating Eq.\,\eqref{eq:eventrateICDC} for each bin), respectively. We note that Ref.\,\cite{IceCube:2016dgk} (Fig.\,6) provides statistical uncertainty in background expectations for IceCube and DeepCore. Assuming the data follow Poisson statistics, we take the statistical uncertainty in the data, $\sigma_{\rm data}^i=\sqrt{N_{\rm data}^i}$. We add the uncertainty in background and the data in quadrature to obtain $\sigma^i$ such that $\sigma^i=\sqrt{(\sigma_{\rm bkg}^i)^2 + (\sigma_{\rm data}^i)^2}$. We have not taken any systematic uncertainty in our analysis, as the IceCube Collaboration has not made that information public in their analysis. We encourage the IceCube Collaboration to publish their full systematic uncertainties in these kinds of searches in their upcoming works.  In our analysis the minimum value of $\chi^2$ ($\chi^2_{\rm min}$) is achieved in the absence of the signal. For a fixed DM mass, we iterate over $\sigma_e$ until $\chi^2-\chi^2_{\rm min} \approx 2.71$. This sets a $95\%$ confidence level constraint on the DM-electron scattering cross section. We performed the same analysis to obtain the constraint from the DeepCore (2016) dataset. 

{\bf DeepCore (2021) \cite{IceCube:2021xzo}:} Unlike the analysis mentioned above, in this case, IceCube Collaboration has focused only on 6.75 years of DeepCore data.  DeepCore has a lower energy threshold and can detect both tracklike and cascadelike signatures. The median angular resolutions for neutrino of energies $10\,$GeV and $200\,$GeV are $\sim 35^{\circ}$ and $\lesssim 5^{\circ}$ respectively. Given that the Sun has $\sim 0.5^{\circ}$ angular diameter in the sky, an analysis like the previous one is difficult to perform. Rather, we compared the annihilation rate, given in Eq.\,\eqref{eq:annrate}, with the same of Ref.\,\cite{IceCube:2021xzo} for each of the considered channels to obtain our results.
\begin{figure*}
\centering
\subfloat[\label{sf:ee}]{\includegraphics[width = \columnwidth]{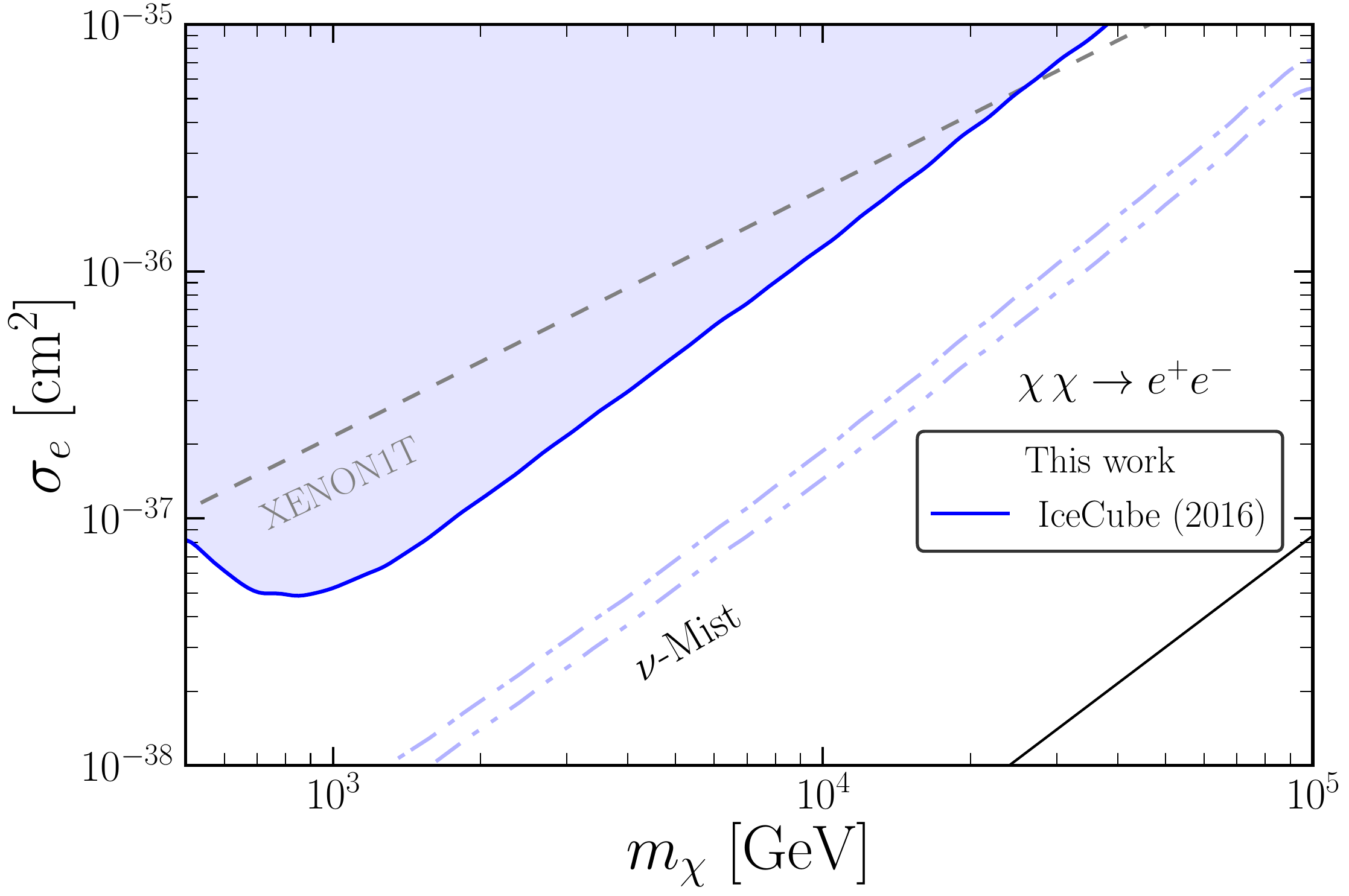}}~~
\subfloat[\label{sf:nunu}]{\includegraphics[width = \columnwidth]{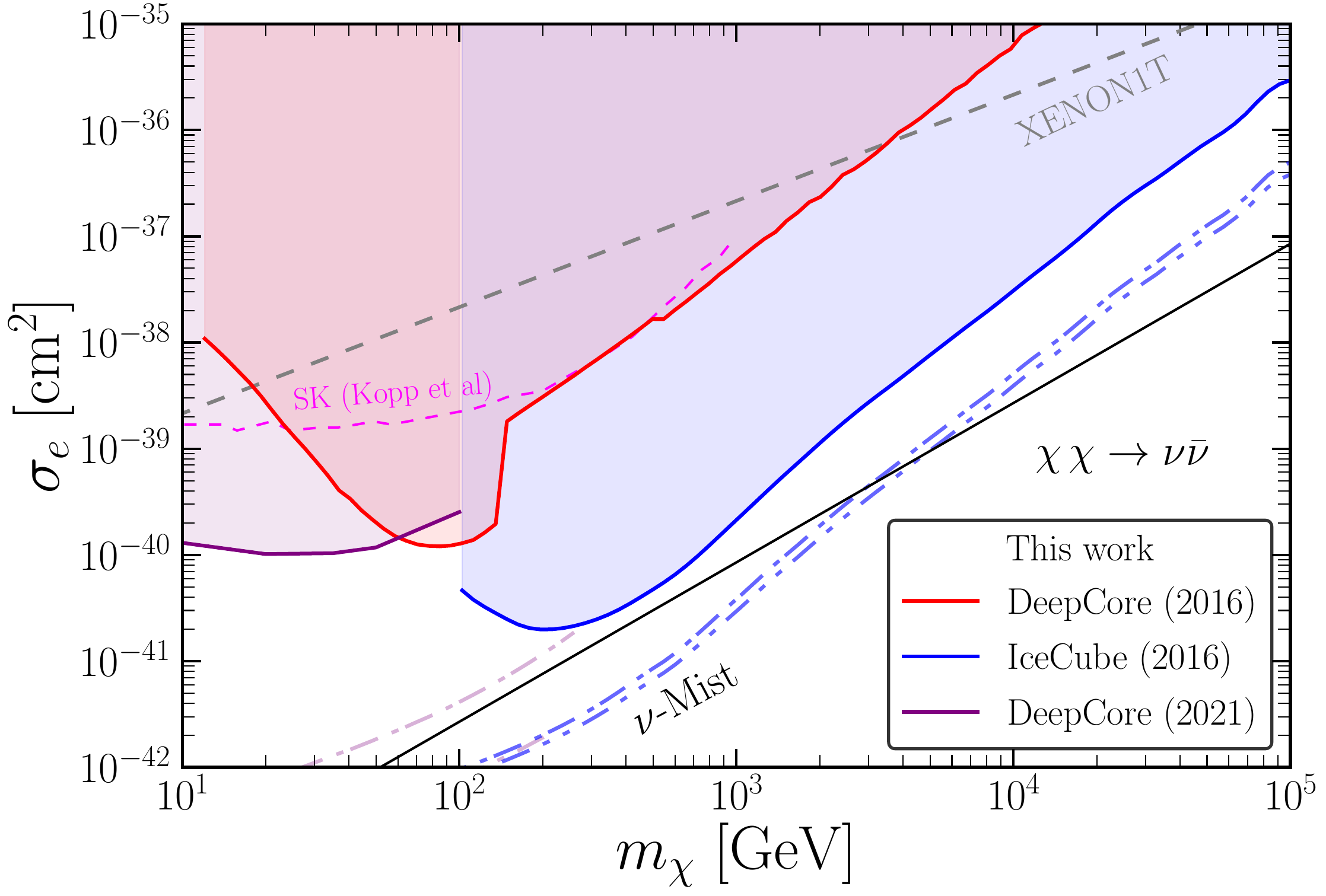} }~~
\caption{Excluded region of the DM-electron scattering cross section for $e^+ e^-$ (left panel) and $\nu \bar{\nu}$ (right panel) final states. Other relevant details are same as the Fig.\,$1$.}
	\label{fig:otherfinalstates}
\end{figure*}
\section{Results and Discussion}
\label{sec:result}
For captured DM annihilation to $\tau^+\tau^-$, our results are shown in Fig.\,\ref{fig:limit}. The region above the gray dashed line is excluded by lab-based direct detection (DD) experiment, obtained from $1/m_{\chi}$ extrapolated XENON1T S2-only limit\,\cite{XENON:2019gfn}. The solid blue, red, and purple lines show the limits that we derive using IceCube (2016), DeepCore (2016), and DeepCore (2021) data, respectively. We also display the previously obtained SK constraint \cite{Kopp:2009et} by the dashed magenta line. It is clear that our constraints are stronger than the previous limits for the DM mass range $\in [ 10,\,10^5]\,$GeV.\footnote{We do not extend our limit above DM mass $100\,$TeV to respect the partial wave unitarity bound\,\cite{Griest:1989wd,Shoemaker:2011vi}.}

IceCube loses sensitivity below DM masses $\sim 200\,$GeV, as the neutrinos produced from such DM remain below the detector's threshold. Similarly, the DeepCore (2016) analysis is sensitive to DM masses $\gtrsim 20\,$GeV. However, for the DeepCore (2021) data, better event reconstruction\,\cite{IceCube:2021xzo,IceCube:2017lak} and use of high-efficiency digital optical modules facilitate the detection of neutrinos with energies $\gtrsim 5\,$GeV. Thus, the DeepCore (2021) limit extends up to DM mass $\sim 10\,$GeV.  Our bounds weaken above DM mass $\sim 300\,$GeV owing to the combined effect of neutrino attenuation inside the Sun and fall of the capture rate with the increment of the DM mass. In a realistic particle physics model, DM may not annihilate to a particular final state with $100\%$ branching ratio (as considered in this work). However, even if the DM annihilation branching ratio to $\tau^+ \tau^-$ or $\nu \bar{\nu}$ is $\sim 10\%$, our bound will be strongest in large regions of the parameter space. In passing we point out that our framework would also be useful to probe strong DM-electron scattering cross section to which underground DD experiments are blind. A detailed study of that including multiple scattering is left for future work. The region below the dot-dashed and double dot-dashed light purple and blue lines indicate ``$\nu$-Mist", where DM induced neutrino events will be the same or less than that of the SA$\nu$\,\cite{Arguelles:2017eao, Ng:2017aur, Edsjo:2017kjk}. We calculate this following the method presented in Ref.\,\cite{Ng:2017aur}.\footnote{In some regions of this parameter space, the DM capture rate would not be in equilibrium for typical annihilation rates, thus we use the full solution\,\cite{Peter:2009mk} to obtain the DM induced neutrino flux.} The dashed and dot-dashed lines correspond to two different models of SA$\nu$.  In this region of the parameter space, with sufficient exposure it would be possible to differentiate DM induced neutrino events from SA$\nu$ by doing a spectrum analysis. 

In Fig.\,\ref{fig:otherfinalstates}, we present our results for the final states $e^+e^-$ and $\nu \bar{\nu}\equiv \frac{1}{3} \left(\nu_e \bar{\nu}_e+\nu_{\mu} \bar{\nu}_{\mu}+\nu_{\tau} \bar{\nu}_{\tau}\right)$. For the $e^+e^-$ channel, the only source of producing neutrinos is the EW correction; thus, we only get limits for DM masses above $500\,$GeV (below this DM mass \texttt{\raisebox{0.82\depth}{$\chi$}aro$\nu$} does not include the small EW correction), and the obtained bound is stronger than the XENON1T DD bound for large regions of DM mass. For the $\nu \bar{\nu}$ final state, our constraints are much stronger than the DD and SK bounds in most of the parameter space. The steep increase in the DeepCore (2016) bound (red solid line in Fig.\,\ref{sf:nunu}) from DM mass $130\,$GeV to $140\,$GeV is due to the absence of the angular resolution beyond neutrino energy $\sim 139\,$GeV in Ref.\,\cite{IceCube:2016dgk} and thus missing peak of the spectrum for rest of the DM masses $\gtrsim 140\,$GeV. These results indicate the exciting prospect of probing DM-electron scattering in future neutrino telescopes like Hyper-Kamiokande\,\cite{Hyper-Kamiokande:2018ofw}, KM3NeT\,\cite{KM3Net:2016zxf}, IceCube Upgrade\,\cite{Ishihara:2019aao}, Baikal-GVD\,\cite{Baikal-GVD:2020xgh}, IceCube Gen2 facility\,\cite{Clark:2021fkg}, etc. 

In Figs.\,\ref{fig:limit} and \ref{sf:nunu} we see that some parts of the $\nu$-Mist fall below the DM thermalization line, which means that for that particular region of parameter space, DM is not fully thermalized inside the Sun. DM annihilation from partially thermalized DM has been studied in Refs.\,\cite{Bell:2023ysh, Acevedo:2024ttq}. We leave a detailed study of partially thermalized DM annihilation in the Sun for future explorations. 
\\

\section{Conclusions}
\label{sec:conclusions}
Low threshold DD experiments are most sensitive to DM-electron scattering for DM masses $\sim 10- 100\,$MeV, but their sensitivity decreases for DM masses $\gtrsim 1\,$GeV due to the decrement in the DM flux. In this paper, we study a novel strategy to probe DM-electron scattering in the DM mass range $10-10^5$ GeV. In our framework, DM gets captured inside the Sun through DM-electron scattering. The captured DM annihilates to produce different SM final states. Primary and secondary neutrinos produced from these SM states can escape the Sun and be detected in IceCube.

We have analyzed IceCube (2016), DeepCore (2016), and DeepCore (2021) data to probe the above scenario. In the absence of any signal, we found the strongest constraint in large regions of DM parameter space. IceCube (2016) data provide the strongest constraint for DM-electron scattering in the DM mass range $200-10^5\,$GeV. DeepCore data give the strongest bounds in the mass range of $\sim 10- 100\,$GeV. DM annihilation final states which produce copious amounts of neutrinos (e.g., $\tau^+\tau^-$ and $\nu \bar{\nu}$), lead to a stronger bound than a neutrino-poor final state (e.g., $e^+e^-$). For  the $\tau^+\tau^-$  and $\nu \bar{\nu}$ final state, our bounds are stronger than the previous bounds in large regions of the parameter space. This suggests that upcoming neutrino telescopes (like Hyper-K, KM3NeT, etc.) would possibly be able to discover the signature of DM-electron scattering. We have only focused on the velocity-independent DM-electron scattering cross section, bounds on velocity-dependent cross section would be considered elsewhere.

\emph{\textbf{Acknowledgments --}} We  thank Carlos Arguelles, Debajit Bose, Celine Boehm, Dibya S. Chattopadhyay, Raghuveer Garani, Joachim Kopp, Subhendra Mohanty, Toby Opferkuch, Nirmal Raj, Tirtha S Ray, Tracy Slatyer, Juri Smirnov, Javier Acevedo, Deep Jyoti Das, and Srijan Kumar for comments and discussions.  We also thank Pierluca Carenza, Tim Linden, Thong Nguyen, and Axel Widmark for detailed discussions on DM capture rate. TNM thanks Rahool Kumar Barman for help with computation. TNM would like to acknowledge the financial support from the Australian Research Council through the ARC Centre of Excellence for Dark Matter Particle Physics and from the Department of Atomic Energy, Government of India, for the Regional Centre for Accelerator-based Particle Physics (RECAPP), Harish Chandra Research Institute. AKS acknowledges the Ministry of Human Resource Development, Government of India, for financial support via the Prime Ministers’ Research Fellowship (PMRF). SM acknowledges the Department of Science and Technology (DST), Government of India, for financial support via the Kishore Vaigyanik Protsahan Yojana (KVPY: now discontinued and merged with the INSPIRE fellowship). RL acknowledges financial support from the Infosys foundation (Bangalore), institute start-up funds, the Department of Science and Technology (Govt. of India) for the grant SRG/2022/001125, and ISRO-IISc STC for the grant
no. ISTC/PHY/RL/499.\\
The first three authors contributed equally to this work.

\appendix 
\section{Appendix A: Dark Matter thermalization inside the Sun}
\label{appendix}

After being captured inside the Sun, DM particles follow an elliptical orbit through the Sun. Meanwhile, the captured DM loses its energy through subsequent scattering with solar electrons. The thermalization process of DM in the Sun can be divided into two parts - first and second thermalization. Here we closely follow the thermalization prescription given in Refs.\,\cite{Kouvaris:2010jy,Acevedo:2019gre,Acevedo:2020gro,Janish:2019nkk}. But note that we are considering DM capture in the Sun due to DM-electron scattering. 
\subsection*{Energy loss rate for DM-electron scattering}
\label{sec:kinematics}
Let us assume that a DM particle of mass $m_{\chi}$ with initial velocity $v_{\chi}'$ scatters off an electron with mass $m_e$ and initial velocity $v'$. The energy loss rate per scatter of a DM particle is (see Appendix B of Ref.\,\cite{Acevedo:2020gro}),
\begin{eqnarray}
\begin{split}
\frac{dE}{dt} & = n\,\sigma_e\,|v'-v_\chi'| \,(\Delta E)\,,\\
& =2n\,\sigma_e\,|v'-v_\chi'|\, (v'-v_{\chi}') \left(v_{\chi}'+\frac{v'-v_{\chi}'}{m_{\chi}}     \mu\right)\mu\,,
\end{split}	
\label{eq:Kinematics1}
\end{eqnarray}
%\begin{eqnarray}
%	\Delta E= 2(v_e-v_{\chi})& &\left(v_{\chi}+\frac{v_e-v_{\chi}}{m_{\chi}}     \mu\right)\mu,
%	\label{eq:Kinematics1}
%\end{eqnarray}
where $\mu = \frac{m_e m_{\chi}}{m_e+m_{\chi}} \sim m_e$ (when $m_e \ll m_\chi$), is the reduced mass of the DM-electron system, $n$ is the number density of target electrons, and $\Delta E$ is the energy transfer in a single scatter. The above equation has been derived assuming 1D scattering kinematics. The effects of full 3D scattering are negligible in our scenario.

Owing to the smaller mass, the solar electrons have a much larger velocity than the solar protons. Besides, after thermalization DM will settle down in a thermal sphere with a thermal velocity much less than the solar electron velocity as, $\sqrt{T_{\rm core}/m_{\chi}} \ll \sqrt{T_{\rm core}/m_e}$, where $T_{\rm core}$ is the temperature at the solar core.

So in our analysis of DM thermalization, we can assume the thermalization to be happening in a `viscous' regime,
\begin{eqnarray}
v'\gg v_{\chi}'.
\end{eqnarray}
Now, we are interested in average energy loss rate, so averaging over all electrons, $\langle v'\rangle=0$. As a result, in Eq.\,(\ref{eq:Kinematics1}) we can drop the terms that change sign under $v'\rightarrow-v'$ thus obtaining,
\begin{equation} \label{Kinematics2}
\begin{split}
\ev{\frac{dE}{dt}}
& = 2n\,\sigma_e\,|v'|\,\left[-m_e v_{\chi}'^2\left(1-\frac{m_e}{m_\chi}\right)+ \frac{m_e^2v'^2}{m_\chi}\right]\,, \\
& \sim n\,\sigma_e\,|v'|\,\left[-m_e v_{\chi}'^2+ \frac{m_e^2v'^2}{m_\chi}\right].
\end{split}
\end{equation}

For our parameters of interest, the first term dominates over the last term. So,
\begin{eqnarray}\label{rate}
\ev{\frac{dE}{dt}} \sim -m_e n\,\sigma_e\,|v'|\,v_\chi'^2.
\end{eqnarray}\\

\subsection*{First thermalization} After one scattering with solar electrons when the DM velocity becomes less than the escape velocity at that point, DM becomes gravitationally bound to the Sun. As the density of the Sun is maximum at the center, we can consider the solar center to be the focus of such DM orbits. 
Now after each scattering with a solar electron, DM loses a part of its initial kinetic energy,
\begin{equation}\label{loss}
\begin{split}
\Delta E & =\frac{2m_e}{m_\chi} E_{\rm kin}\\
& = \frac{2m_e}{m_\chi}(E-m_\chi\Phi (r))\,,	
\end{split}
\end{equation}
where,
\begin{eqnarray}
\Phi (r)= -\frac{G M_\odot}{R_\odot}+ \int_{R_\odot}^{r} \frac{G M(r')}{r'^2} \,dr',
\end{eqnarray}
where, $E_{\rm kin},E,$ and    $m_{\chi}\Phi$ are the kinetic energy, total energy, and potential energy of the DM particle, respectively. Also, $M(r)=\int_{0}^{r}4\pi r'^2\rho_\odot(r')\,dr'$ is the mass enclosed in a sphere of radius $r$, with $
\rho_\odot$ being the solar density.

For obtaining the average energy loss we can integrate Eq.\,(\ref{loss}) over the size of the Sun,
\begin{eqnarray}
\langle \Delta E\rangle = \frac{1}{R_\odot}\int_{0}^{R_\odot}2m_e\left(\frac{E}{m_\chi}-\Phi(r)\right)\,dr.
\end{eqnarray}
During the first thermalization phase, each DM particle will cross the solar body twice. Treating the star as a point object for simplicity, the orbital period of DM during this phase becomes,
\begin{eqnarray}
\Delta t=2\pi \sqrt{\frac{\left(a_p(E)\right)^3}{G M_\odot}}\,,
\end{eqnarray}
where, $a_p(E)$ is the semimajor axis distance of the captured DM orbit given by,
\begin{eqnarray}
a_p(E)=-\frac{GM_\odot m_\chi}{E}.
\end{eqnarray}

Therefore the average energy loss rate becomes,
\begin{eqnarray}\label{enloss}
\frac{dE}{dt} = -\frac{\langle \Delta E\rangle}{\Delta t}\times\tau.\,
\end{eqnarray}
where $\tau$ is the optical depth for the DM-electron scattering in the Sun, given by,
\begin{equation}
\begin{split}
\tau & = \int_{0}^{R_\odot}n (r)\,\sigma_e\,dr\,,\\
& \sim  \frac{\sigma_e}{10^{-36}\,\rm cm^{2}}\,,
\end{split}
\end{equation}
where $n$ is the number density of the solar electrons.
\begin{figure}
	\centering
	\includegraphics[scale=0.245]{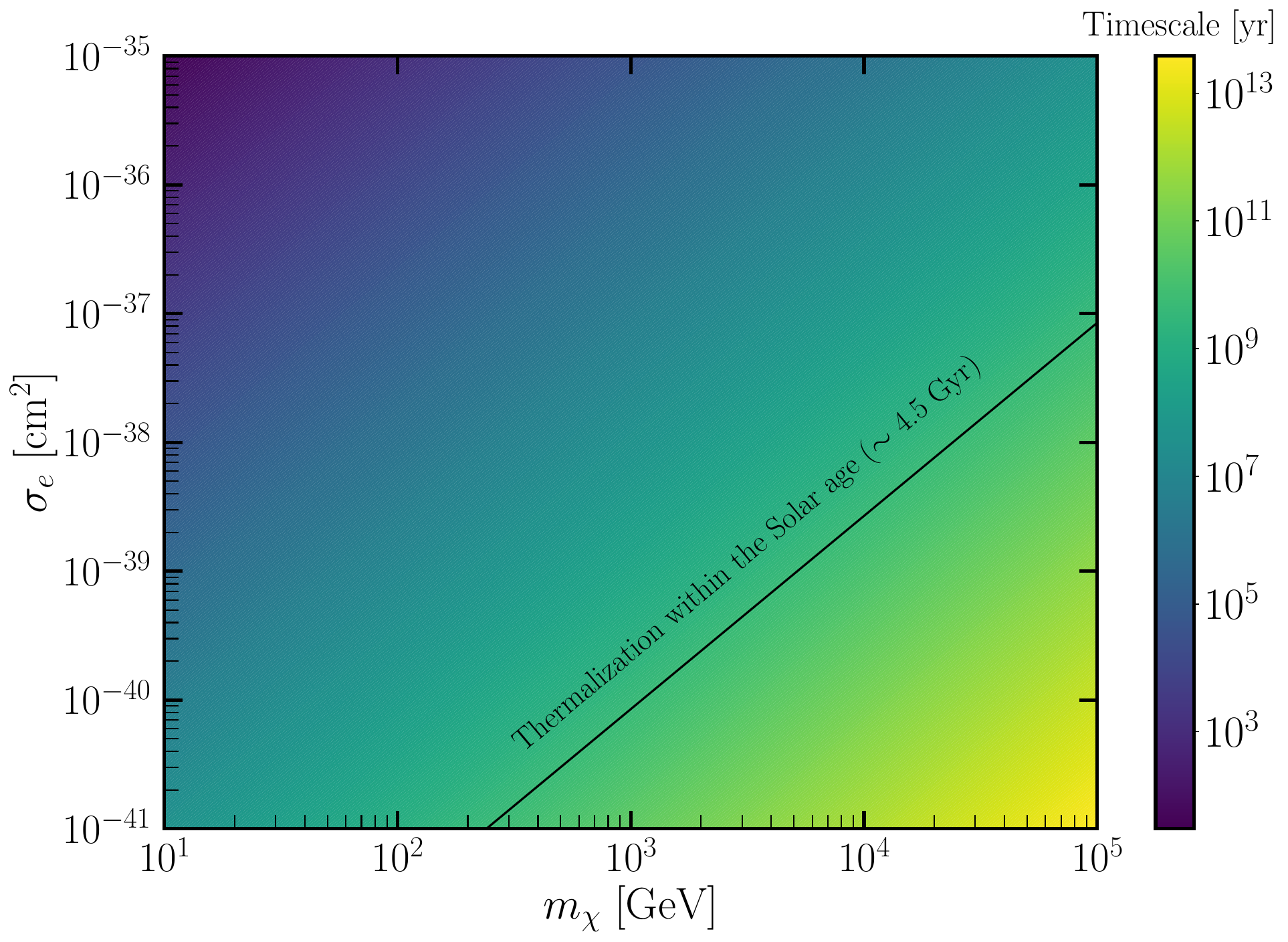}
	\caption{Thermalization timescales for DM-electron scattering inside the Sun. The black solid line shows  DM-electron cross sections above which DM is thermalized within the lifetime of the Sun.}
	\label{fig:thermalplot} 
\end{figure}
Let us now define the variable, $\epsilon=R_\odot/a_p$, which is the ratio between the radius of the Sun and the semimajor axis of the transiting DM particles. With this substitution Eq.\,(\ref{enloss}) becomes,
\begin{eqnarray}\label{enloss2}
\frac{d\epsilon}{dt} = -\frac{\langle \Delta \epsilon\rangle}{\Delta t}\times\tau,
\end{eqnarray}
with,
\begin{eqnarray}
\langle \Delta \epsilon\rangle = \frac{1}{R_\odot}\int_{0}^{R_\odot}2m_e\left(\epsilon-\frac{R_\odot\Phi(r)}{GM_\odot}\right)\,dr.
\end{eqnarray}
Now Eq.\,(\ref{enloss2}) has to be integrated from the initial value of $\epsilon$, just after capture, to the final value when first thermalization ends. Let us first focus on the upper limit of the integration.

By definition, first thermalization ends when the DM orbit completely enters the Sun, thus $\epsilon_f\sim R_\odot/R_\odot=1$. On the other hand numerically, $R_\odot/a_p\sim 2m_e/m_\chi$. One way to understand this is that the right-hand side is the energy loss per scatter. As we increase the number of scatters, the semimajor axis of the DM orbit $a_p$ becomes smaller and eventually the DM orbit enters the Sun completely. So our initial limit of integration becomes, $\epsilon_i=2m_e/m_\chi$.
Finally, after integrating Eq.\,(\ref{enloss2})  we obtain the time required for first thermalization to take place,
\begin{equation}
\begin{split}
t^{\mathrm I} & =\frac{\pi R_\odot^\frac{3}{2}\,m_\chi}{\tau\, m_e \sqrt{G\,M_\odot}}
\int_{\frac{2m_e}{m_\chi}}^{1}\frac{d\epsilon}{\epsilon^\frac{3}{2}(2.38-\epsilon)}\\
& \simeq \, 3.8\times 10^5 \,\,{\rm yr}\,\left(\frac{m_\chi}{10^3 \,\rm GeV}\right)^{\frac{3}{2}}\,\left(\frac{10^{-36} \,\rm cm^2}{\sigma_e}\right).
\end{split}
\end{equation}

\subsection*{Second thermalization}
After the DM orbits shrink inside the Sun, DM keeps on scattering with the solar electrons further and loses its energy to finally settle down inside the solar core with a thermal velocity $\sim \sqrt{3T_{\rm core}/m_\chi}$.
It is important to remember that we are still working in the viscous regime where $v_\chi \ll v_e$.
The energy loss rate per scatter is then given by Eq.\,(\ref{rate}) ,
\begin{equation}\label{thermal2}
\begin{split}
\frac{dE}{dt} & \sim - m_e n\,\sigma_e\,v'\,v_\chi'^2\\
& \sim -\frac{2 n\,m_e\,\sigma_e\,v'}{m_\chi}\left(\frac{1}{2}m_\chi\, v_\chi'^2\right)\\
& \sim -\frac{2 n\,m_e\,\sigma_e\,v'}{m_\chi}\,(E-\Phi(0)),
\end{split}	
\end{equation} 
where, $\Phi(0)$ is the gravitational potential of the Sun at minima. Integrating Eq.\,(\ref{thermal2}) from initial energy $E_i$ to final energy $E_f$ one obtains,
\begin{equation}
t^{\mathrm{II}}=\frac{m_\chi}{2 n\,m_e\,\sigma_e\,v_e} \log\left[\frac{E_i-\Phi(0)}{E_f-\Phi(0)}\right].
\end{equation}
After completion of first thermalization, a DM has a kinetic energy of $\sim\frac{G M_\odot m_\chi}{R_\odot}$, so, $E_i=\Phi(0)+\frac{G M_\odot m_\chi}{R_\odot}$. Final velocity of DM particles after thermalization will be $\sim \sqrt{\frac{3T_{\rm core}}{m_\chi}}$, thus, $E_f=\Phi(0)+\frac{3T_{\rm core}}{2}$.
Thus, the second thermalization timescale is,
\begin{equation}
%\begin{split}
%	t^{\rom{2}} & =\frac{m_\chi}{2 n_e\,m_e\,\sigma_e\,v_e} \log\left[\frac{\frac{GM_\odot m_\chi}%{R_\odot}}{\frac{3T_{\rm core}}{2}}\right],\\&
t^{\mathrm{II}}	\sim \left(\frac{2\times10^{-40} \,\rm cm^2}{\sigma_e}\right)\left(\frac{m_\chi}{\rm GeV}\right)\left(0.06+\log\left(\frac{m_\chi}{\rm GeV}\right)\right)\,\,\rm yr.	
%\end{split}
\end{equation}
Hence the total thermalization timescale is given by,
\begin{eqnarray}
t=t^{\mathrm{I}}+t^{\mathrm{II}}\,.
\end{eqnarray}
Our result for the thermalization timescale, considering DM-electron scattering is shown in Fig.\,\ref{fig:thermalplot}. For our parameters of interest we find that, within the age of the Sun, DM particles will indeed be thermalized due to DM-electron scattering.   
%%%%%%%%%
\section{Appendix B: Equilibrium timescale} 
\label{app:appB}
\begin{figure}
	\centering
	\includegraphics[width=\columnwidth]{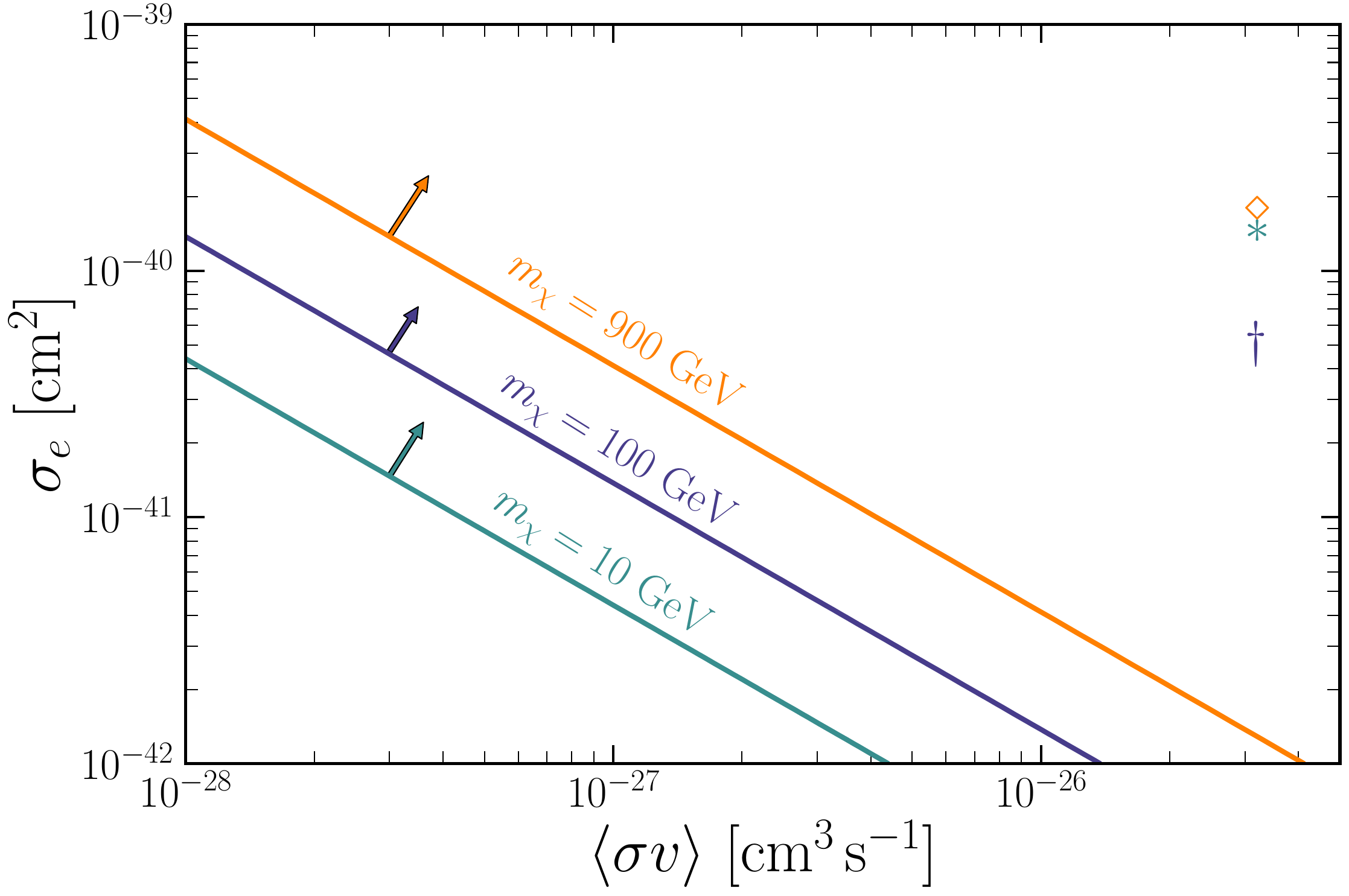}
	\caption{The different colored solid lines indicate the contours above (on) which captured DM equilibrium timescale will be smaller than (equal to) the solar age for different DM masses, as mentioned in the figure. For completeness we have also shown our bounds assuming annihilation cross section $\langle \sigma v \rangle =  3 \times 10^{-26}$cm$^3$s$^{-1}$ to $\nu \bar{\nu}$ by the star, dagger, and diamond for DM masses $10\,$GeV, $100\,$GeV, and $900\,$GeV respectively. }
	\label{fig:eqtime} 
\end{figure}

%%%%%%%%
In Fig.\,\ref{fig:eqtime}, we present the contours where the captured DM equilibrium timescale is equal to solar age. The teal, dark blue, orange color lines are for DM masses $10\,$GeV, $100\,$GeV, and $900\,$GeV respectively. Above these lines (indicated by the arrows) the captured DM equilibrium timescale would be less than the solar age. We also show bounds obtained with captured DM annihilation to $\nu \bar{\nu}$ in this work for aforesaid DM masses by the star, dagger, and diamond for a typical choice of thermal relic cross section. Clearly even for a smaller choice of thermally averaged annihilation cross section, DM would also equilibrate inside the Sun for the achieved DM-electron scattering cross section. For the $e^+ e^-$ final state, at DM masses above $500\,$GeV, none of the current constraints approach the thermal relic cross section \cite{John:2021ugy, Leane:2018kjk, Bergstrom:2013jra}. For annihilation into $\tau^+ \tau^-$, the strongest constraint within our mass range occurs at a DM mass of $10\,$GeV, with a bound of $4.30 \times 10^{-27} \,\rm{cm}^3\,\rm{s}^{-1}$ \cite{McDaniel:2023bju}. As shown in Fig.\,\ref{fig:eqtime}, even if the DM annihilation cross section is taken at the current bound probed by Fermi data, the capture and annihilation rates will be in equilibrium for the DM-electron scattering cross section range that IceCube is probing. For the $\nu \bar{\nu}$ final state, current limits in our mass range of interest are approximately 1 order of magnitude weaker than the thermal relic cross section \cite{Arguelles:2019ouk}. Thus, our results are consistent with indirect detection limits on the thermally averaged cross section. 

%
%
%
%

%%%%%

\bibliographystyle{JHEP}
\bibliography{ref.bib}

\end{document}